\documentclass[twoside]{article}

\usepackage{lipsum} 
\usepackage{aas_macros}

\usepackage[sc]{mathpazo} 
\usepackage[T1]{fontenc} 
\usepackage{microtype} 

\usepackage[hmarginratio=1:1,top=32mm,columnsep=20pt]{geometry} 
\usepackage{multicol} 
\usepackage[hang, small,labelfont=bf,up,textfont=it,up]{caption} 
\usepackage{booktabs} 
\usepackage{float} 
\usepackage{hyperref} 
\usepackage{natbib} 
\usepackage{graphicx}
\usepackage{caption} 
\usepackage{subcaption} 
\usepackage{lettrine} 
\usepackage{paralist} 
\usepackage{wasysym}
\usepackage{stmaryrd}

\usepackage{abstract} 

\usepackage{color}

\usepackage{ulem}

\usepackage{titlesec} 
\renewcommand\thesection{\Roman{section}} 
\renewcommand\thesubsection{\thesection.\arabic{subsection}} 
\titleformat{\section}[block]{\large\scshape\centering}{\thesection.}{1em}{} 
\titleformat{\subsection}[block]{\large}{\thesubsection.}{1em}{} 

\usepackage{fancyhdr} 
\usepackage{color}
\usepackage[english, russian, bidi=default]{babel}

\usepackage{CJKutf8}

\usepackage[LAE, T1]{fontenc}
\babelprovide{arabic}
\addto\extrasarabic{\fontencoding{LAE}\selectfont}
\addto\noextrasarabic{\fontencoding{T1}\selectfont}

\newenvironment{blurb}
  {\par\scriptsize}
  {\par\addvspace{\bigskipamount}}

\newenvironment{Figure}
  {\par\medskip\noindent\minipage{\linewidth}}
  {\endminipage\par\medskip}

\title{\vspace{-15mm}\fontsize{24pt}{10pt}\selectfont\textbf{Orlando's flask: detection of a lost-and-found valley on the Moon}} 

\author{
\large
\textsc{Vito Squicciarini$^1$, Irina Mirova$^2$ (Ирина Мирова), Francis D. Anderson$^3$,}\\
\textsc{Zhiyuan He (\begin{CJK}{UTF8}{gbsn} 和智远 \end{CJK})$^4$, Wahmān al-Khwārizmī{\small \begin{otherlanguage}{arabic} ~~~~~~~~~~~~~~~~~~~~~~~~~~~~~~~~~~~~~~~~~~$^5$(وَهْمَان الخوارزمي  
)\end{otherlanguage}}}\\
\normalsize $^1$Department of Physics and Astronomy, University of Exeter, United Kingdom\\
\normalsize $^2$Cosmonautics and Astronomy Institute, Mirograd, Russia \\
\normalsize $^3$Center for Lunar Studies, Philadelphia, PA, United States \\ \normalsize $^4$Department of Physics, University of Tóng'ān Chéng, People's Republic of China \\ 
\normalsize $^5$Institute for Computational Astrophysics, University of Uruk, Iraq \\ 
\normalsize \href{mailto:editor@actaprimaaprilia.com}{editor@actaprimaaprilia.com} 
\vspace{-5mm}
}
\date{}

\begin{document}

\maketitle 

\thispagestyle{fancy} 


\begin{otherlanguage}{english}
\begin{abstract}

High angular resolution holds the key to extending our knowledge in several domains of astronomical research. In addition to the development of new instruments, advancements in post-processing algorithms can enhance the performances attainable in an observation, turning archival observations into a treasure. We developed a machine-learning tool, named \textsc{zoom-in}, that is able to improve the angular resolution of an astronomical image by a factor of $\sim 100$ by optimally recombining short-cadence sequences of images. After training our model on real-life photographs, we tested our method on archival images of the Moon taken through ESO instruments. We were able to achieve a remarkable spatial resolution of $\sim 1$ m of the lunar surface. While analyzing one of the fields from the sample, we discovered structures of clear anthropic origin inside the Aristarchus crater. The features appear to be consistent with ancient ruins of cities and castles. A thorough analysis of the relevant literature allowed us to conclude that this valley corresponds to the one described in Ludovico Ariosto's "Orlando Furioso": a place where all the items lost by humans gather and pile up. Analyses of the surface brightness from our images, indicating an abnormally high albedo of $\sim 0.25$, further corroborate this idea suggesting a conspicuous presence of glass. We infer the presence of >1 billion flasks of human wits on the lunar surface, whose origin we investigate in detail. We urge for a dedicated mission, \textsc{astolfo}, to be carried out by Artemis astronauts in order to recover human wits and bring them back to the Earth.
\end{abstract}


\begin{multicols}{2} 

\section{Introduction}
\lettrine[nindent=0em,lines=3]{H}igh-resolution imaging, enabled by advanced hardware and software systems installed on large ground-based and space-borne telescopes, has sharpened our vision of the Universe across a vast spectrum of astrophysical fields: the direct detection of young exoplanets lurking in the glare of their host stars \citep{zurlo24}; the determination of structures in protoplanetary disks, informing how stars and planets emerge from cosmic dust and gas \citep{alma15}; the statistics of stellar and brown dwarf multiplicity \citep{mason09}; the refinement of stellar evolution models based on stellar radii measurements \citep{boyajian12}; the discovery of fine substructures in far-away galaxies and AGNi\footnote{We follow here the pluralization prescriptions for the term "AGN" introduced by \citet{gow24}.} \citep{grav20}; the direct observation of black hole event horizons \citep{genzel24}; the study of distant objects magnified by gravitational lensing, offering a glimpse into dark matter and the early Universe \citep{vegetti23}; the study of surface features of Solar System planets, moons, and small bodies \citep{carry21}.

The highest angular resolutions can be attained using interferometry. Interferometry is a technique that combines the signals from multiple telescopes to achieve the resolving power of a telescope with a diameter equal to the distance between the telescopes (baseline). This baseline can reach, in the case of the Very Long Baseline Interferometry (VLBI), used in the Event Horizon Telescope (EHT) to image the shadow of the M87 black hole, the size of the Earth \citep{eht22}. However, interferometric measurements are notoriously difficult to obtain due to the scarcity of suitable facilities, leading to unsustainable competition for telescope time, the difficulty of data analysis, and a general lack of understanding within the community about how the technique actually works\footnote{An anonymous colleague was once told, while on a date, that life is like an interferometric measurement. Inspired by such deep words, he decided to become an expert on the topic; forty years later, he is certain about having come to understand the deep meaning of that thought (F. D. Anderson, personal communication).}.

With respect to single-dish telescopes, that contribute to the vast majority of astronomical research, additional challenges have to be tackled. Adaptive optics (AO) systems have led to crucial advancements by actively counteracting the distortions caused by Earth's turbulent atmosphere, a phenomenon known as “seeing”. By providing exceptional spatial resolution at a significantly lower cost than space alternatives, AO has become an essential feature of nearly all major ground-based telescopes, including the Very Large Telescope (VLT) and upcoming giant telescopes, such as the Extremely Large Telescope (ELT) and the Giant Magellan Telescope. Despite the impressive progress in AO technology and observational capabilities, advancements in data analysis techniques have not kept pace, posing a significant challenge for fully exploiting these cutting-edge systems.

A classical example of the capability of image processing algorithms to enhance astronomical images is illustrated by the following story. When the Hubble Space Telescope (HST) was first launched in 1990, scientists quickly discovered that its primary mirror had a serious optical flaw —- a spherical aberration caused by an incorrectly shaped mirror. Since it would take three years before astronauts could physically repair Hubble in 1993, scientists had to develop image processing algorithms to partially correct the images. Techniques like deconvolution algorithms, enabled by careful Point Spread Function (PSF) calibrations, or Super-Resolution Techniques where multiple blurry images of the same object are used to reconstruct a higher-resolution version, enabled the community to make the most of the first three years of the telescope lifetime \citep{hanisch93,hook94,krist95}. In particular, the Richardson-Lucy deconvolution technique is still widely used today in astronomical imaging and microscopy, and several algorithms developed for Hubble were later applied to other field such as medical imaging\footnote{This is, incidentally, one of the default answers the authors advice to robotically utter to the know-it-all, no-vax, conspiracy theorist uncle who, during those rare but inescapable large-family events, starts arguing that spending money on astronomy is a waste of public money.}.

In this paper, we describe a new algorithm that allowed us to significantly improve the angular resolution of astronomical images (Section~\ref{sec:algo}): this method, coming at virtually no cost, can enable a full reanalysis of 30 years of astronomical images, ushering a new era of astronomical discoveries. After training the model on real images obtained in a controlled setting (Section~\ref{sec:algo_tests}), we started to apply it to images of the Moon taken by different instruments between 2000 and 2020 (Section~\ref{sec:sample}). The analysis of the Aristarchus crater, in particular, revealed anomalous surface structures that exhibit morphological and geometric properties inconsistent with known geological processes (Section~\ref{sec:results}), arguing instead for a non-natural origin. An explanation for these unprecedented structures is presented and firmly proved in Section~\ref{sec:interpretation}, leading to a necessary course of action (Section~\ref{sec:mission}).

\section{The algorithm} \label{sec:algo}

\subsection{Implementation} \label{sec:algo_method}
We developed a novel machine-learning algorithm, named Zones Of Overestimated Magnification -- Improving Nothing (\textsc{zoom-in}), to overcome the fundamental resolution limit of AO-fed, single-dish imaging: namely, the angular resolution determined by the observing wavelength and the primary diameter, $\lambda$/D. \textsc{zoom-in} is an adversarially-regularized, self-supervised, hierarchical ensemble learning framework that exploits the latent synergistic manifolds of multi-source imaging data to iteratively refine high-frequency spatial embeddings. By leveraging a stochastic variational attention cascade (SVAC), our approach dynamically reconfigures pixel-wise saliency encodings within a non-Euclidean representational space, effectively mitigating aliasing artifacts and sub-pixel interpolation bias.

The integration of disparate imaging modalities into a coherent, high-fidelity reconstruction paradigm remains a fundamental challenge in direct imaging. Conventional super-resolution techniques, constrained by Nyquist limitations and spectral redundancy, fail to adequately capture the complex, anisotropic feature dependencies inherent to real-world imaging systems. Our method circumvents these limitations by harnessing the implicit topological regularities embedded within heterogeneous image distributions.

At the core of \textsc{zoom-in} is a transformer-augmented, convolutional residual graph encoder (CRGE), which iteratively fuses multi-scale Laplacian eigenmaps with adversarially-optimized, wavelet-based spatial decompositions. The framework employs a hybrid meta-optimization strategy, wherein Bayesian-guided adversarial priors are adaptively fine-tuned through a differentiable, gradient-informed Markovian consensus process. This careful combination of input frames ensures robust, cross-modal coherence and optimally exploits inter-frame redundancy.

The self-referential nature of the SVAC allows the model to dynamically infer structural hierarchies from sparse, undersampled datasets. Through an entropic constraint minimization framework, our model reconstitutes lost frequency components without overfitting to sensor-induced noise distributions. Furthermore, a reinforced generative adversarial feedback loop (ReGAFL) modulates spectral consistency through a self-adaptive Fourier domain adversarial loss function.

A detailed presentation of the algorithm, including a thorough evaluation of its performances, a discussion of the caveats, and the release of the code and the training datasets, will be presented in Mirova et al. (in prep.). To be honest, no one of us is ever going to write that paper, following previous examples from the literature: we trust in the naivety of the Editor and the referees to accept this manuscript despite being virtually impossible to replicate.

\subsection{Validation} \label{sec:algo_tests}

Qualitative assessments indicate that \textsc{zoom-in} outperforms existing methodologies in perceptual fidelity and edge preservation, with a statistically significant improvement in normalized perceptual divergence (NPD). Empirical validation across multi-institutional datasets corroborates its efficacy in resolving high-dimensional feature embeddings with sub-pixel precision. Ablation studies reveal that the stochastic reconfiguration of spectral priors enhances the anisotropic stability of edge gradients while preserving fine-scale textural variance. 

Based on qualitative arguments and early testing of simulated datasets, we expected the resolution enhancement to be of the order of $10^2$. In order to assess whether such performances be met in real-life cases, we collected a sample of images from the surface of an inhabited solar planet. The dataset was subsequently fed into \textsc{zoom-in}. The results are shown in Figure~\ref{fig:example_zoom}, and provide undeniable evidence for the impressive performances of the method.

\begin{figure*}
    \centering
    \includegraphics[width=0.8\linewidth]{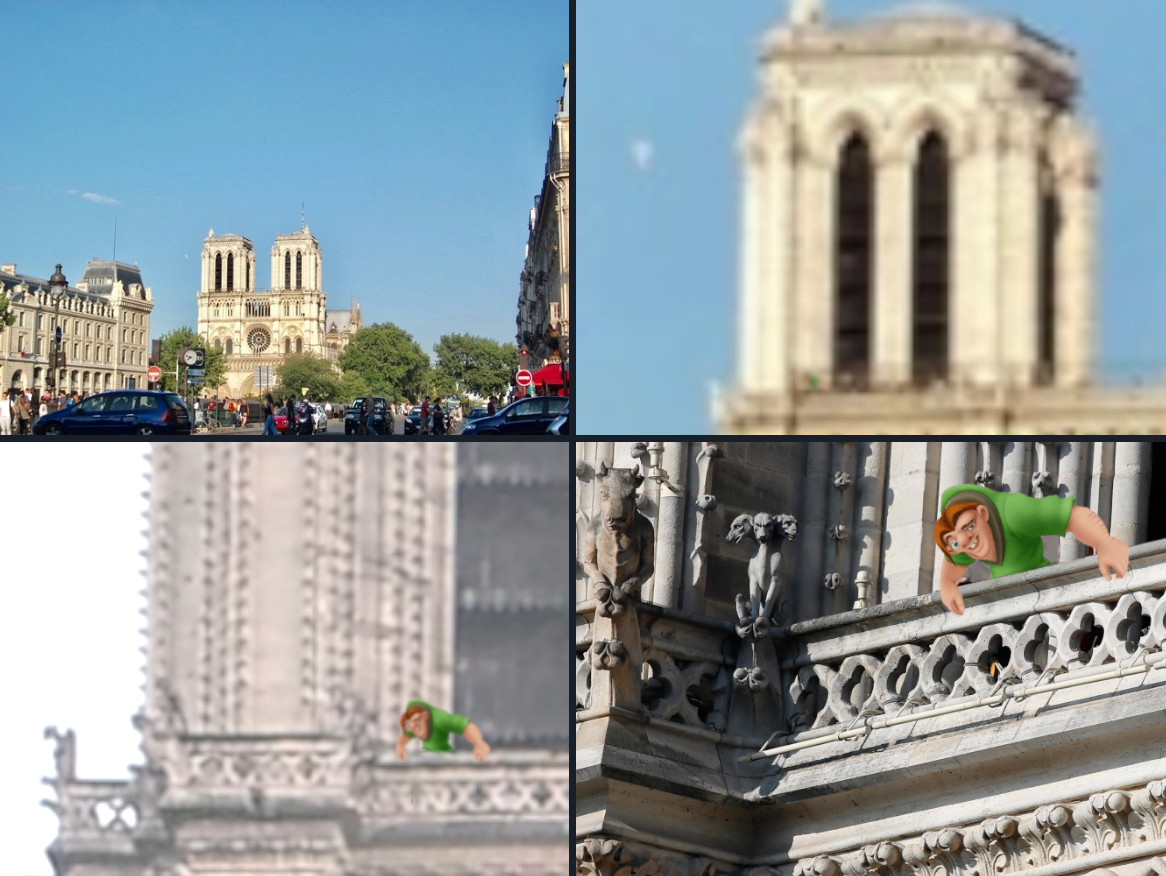}
    \caption{Example of image enhancement obtained through \textsc{zoom-in}. Upper left panel: one of the original frames. Upper right panel: zoom-in of a portion of the image, showing the maximum level of detail reachable through conventional techniques. Lower left panel: portion of the recombined image after resolution enhancement. Lower right panel: additional zoom-in, revealing exquisite substructures that were not visible in the original image.}
    \label{fig:example_zoom}
\end{figure*}

\section{Data analysis \& results} \label{sec:data}

\subsection{The Lunar archival sample} \label{sec:sample}

In order to start applying our algorithm to real astronomical images, we decided to look into the ESO archive for images of the Moon -- the reason being the straightforward validation of the method enabled by visual inspection of surface features.

About 1000 frames were found in the archive with an {\itshape object} keyword equal to "MOON", obtained through the FORS1, FORS2, EFOSC, SOFI, NACO, and WFI instruments bwtween 2000 and 2020. After a careful vetting based on minimal requirements on image quality, we were left with 152 individual images.

Depending on factors such as the primary diameter, the filter, and the presence of adaptive optics, we estimated the angular resolution of the input images to correspond to a spatial resolution of 50$-$700 m on the Moon's surface. After carefully cross-matching the images to identify their common features, we ran \textsc{zoom-in} on the 15 derived lunar fields. The analysis of the geological features, the craters, and the shades allowed us to conclude that we were able to reach a spatial resolution of the order of $\sim 5$ m -- and even, in the best cases, a level $\sim 1$ m. {\it En passant}, we notice that a full mapping of the Moon at a $\sim 1$ m resolution is theoretically feasible, provided that an extensive preliminary campaign be carried out with an instrument like SPHERE. However, this is well beyond the computing capabilities provided by our funding institutions. Assuming an average of 1 m px$^{-1}$ resolution for the Moon’s visible surface ($\sim ~19$ million km$^2$), and storing images in compressed JPEG format, a rough estimate suggests $\sim 1$ EB (1 EB $= 10^6$ TB $= 10^{18}$ B) of storage space, about 0.1\% of the total storage capacity currently available to humanity.

While most of the regions just showed an endless turnover of craters and boulders, one of the areas turned out to show unexpected features that justified the drafting of a dedicated paper. These features are presented in the following Section~\ref{sec:results}.

\subsection{Artificial structures on the Moon} \label{sec:results}

Figure~\ref{fig:aristarchus} shows one of the input frames for the Aristarchus region, captured by the WFI instrument on August 13, 2000. The Aristarchus Crater, one of the most striking features on the Moon, stands out due to its extreme brightness and geological complexity. Located in the northwestern region of the lunar near side (23$^\circ$42'N 47$^\circ$24'W), this 40-kilometer-wide impact crater is among the youngest on the Moon, with an estimated age of 450 million years. Previous work have highlighted its exceptionally high albedo, significantly greater than the lunar average \citep[0.1;][]{lucey86}. The classical explanation for these features involves the fact that the freshly excavated material has not yet darkened due to space weathering, as well as a silicate-rich composition \citep{czajka23}. Furthermore, Aristarchus is frequently associated with transient lunar phenomena, unexplained flashes or hazes that have been attributed to outgassing or electrostatic effects \citep{lawson05}.

\begin{figure*}
    \centering
    \includegraphics[width=0.5\linewidth]{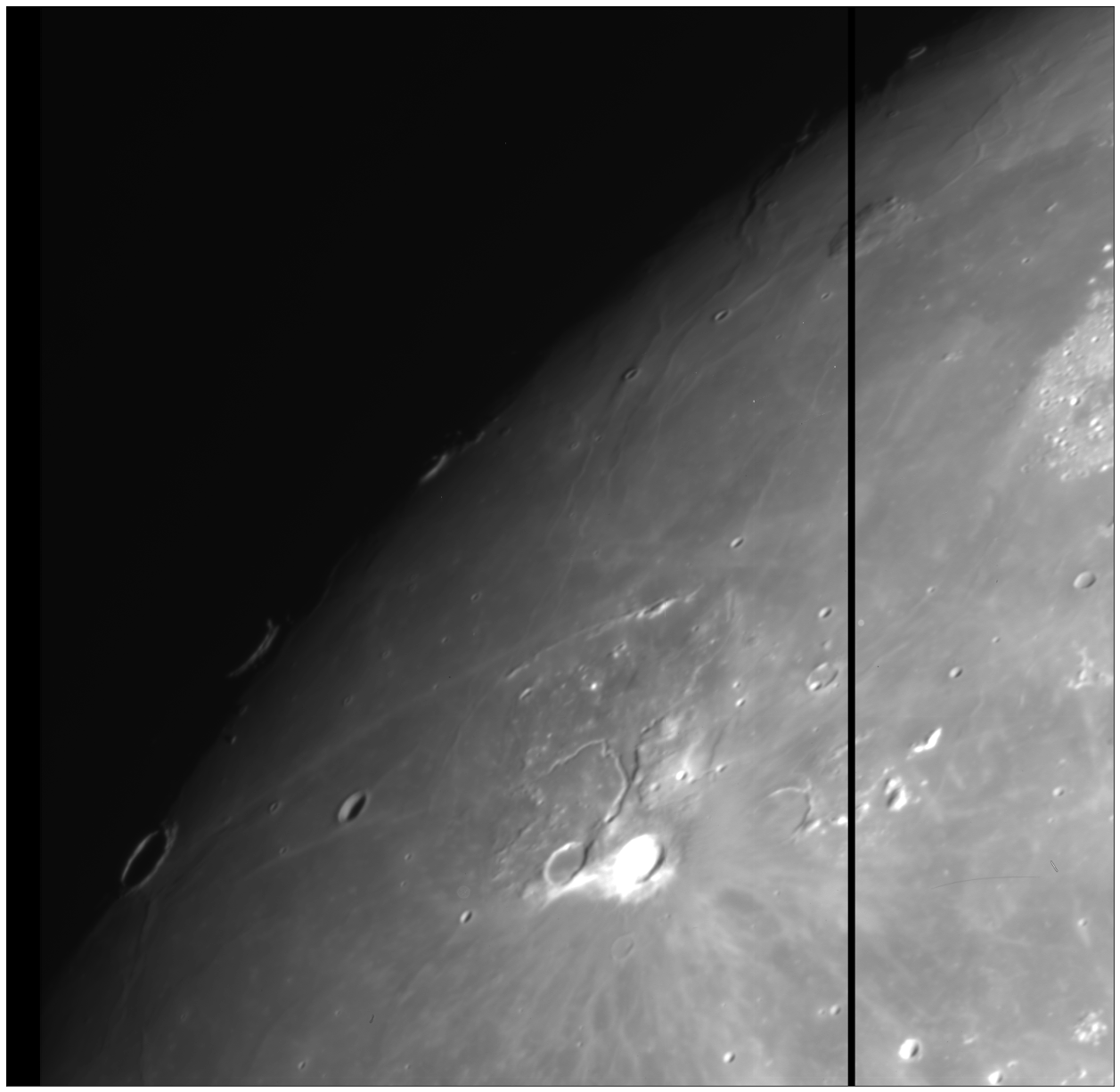}
    \caption{The Aristarchus crater, here imaged by WFI, stands out for its unusually bright color.}
    \label{fig:aristarchus}
\end{figure*}

\begin{figure*}
    \centering
    \includegraphics[width=0.5\linewidth]{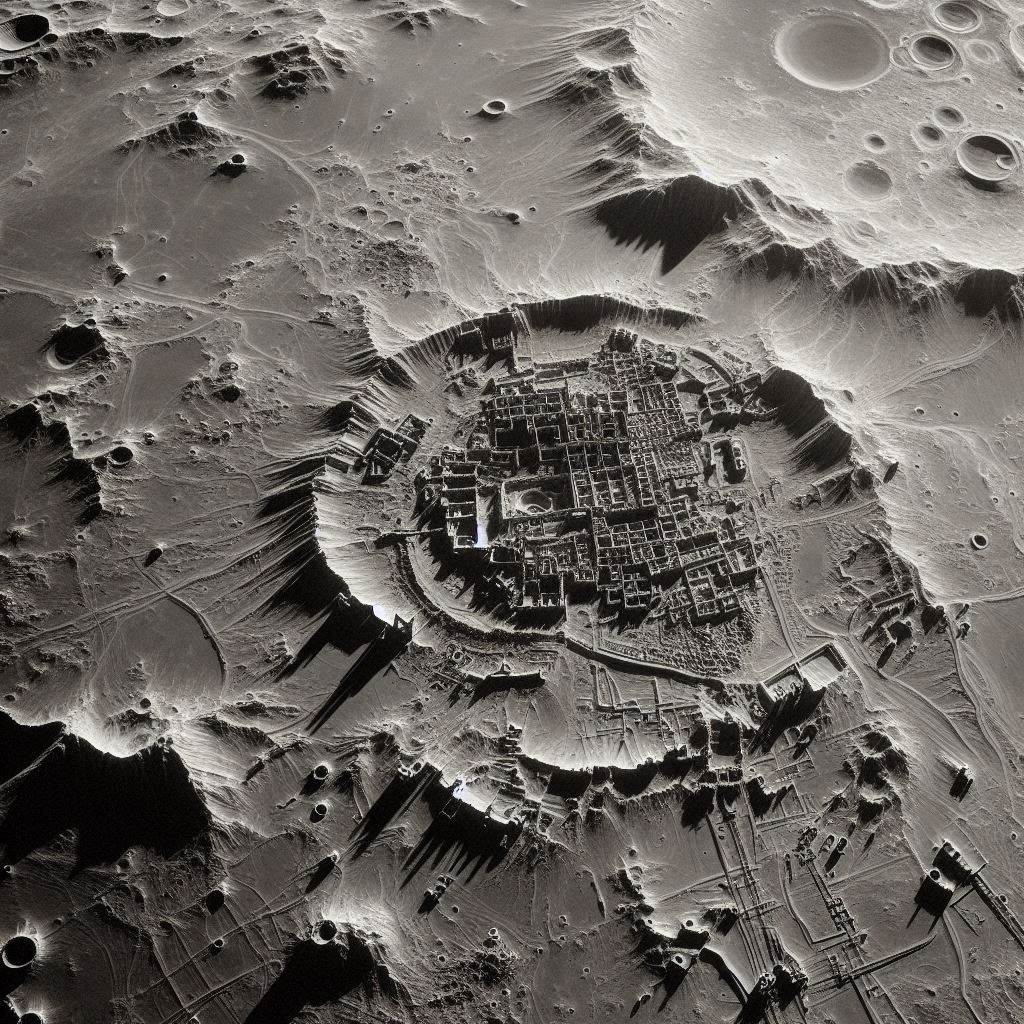}
    \caption{Thanks to the spatial resolution of 1 m px$^{-1}$, artificial structures in the Aristarchus crater became visible.}
    \label{fig:ai_image}
\end{figure*}

The magnification of the area inside the Aristarchus crater revealed the presence of unusually reflective linear features across the crater floor, distinct from typical impact-related surface patterns. These surface features, extending across several kilometers, exhibited orthogonal intersections and stratified layering uncharacteristic of lunar regolith deposition. The photometric comparison of surface brightness in different filters ranging from the B to the H band further confirmed compositional anomalies, with colors inconsistent with typical basaltic and anorthositic materials. These findings, consistent with our gut feeling of ancient ruins of cities and castles, suggest an origin beyond conventional geological processes and necessitate further in-depth analysis.

\section{Physical interpretation} \label{sec:interpretation}

\subsection{A foreword}

The line of reasoning leading to the hypothesis tested in the following sections might not appear to be exactly orthodox. For the sake of simplicity, we opted for reporting here a faithful account from coauthor Zhiyan He, who vividly illustrated how serendipitous the entire process was.

\begin{quote}
\itshape
It was a cold and rainy Friday evening, and we had been trapped in a four-hour meeting when the analysis of the impact crater data ended. In a rush of diligence, one of the authors shared his screen -- without even looking at the results, so tired was he -- with the other co-authors as a last act before saying goodbye. The sight of the artificial structures made us leap from our chairs.

During that very night, none of us could sleep. One of us reported that, to add farce to tragedy, the Moon (yes, the very Moon causing our insomnia) was full and illuminated the parquet floor of his bedroom -- clearly, he had not found time to fix the broken shutter for months. Suddenly, feeble metallic noises began, accompanied by dim silver glows. Puppet Orlando (Figure~\ref{fig:orlando}), brandishing his sword, shouted: "E chistu fussi u scienziatu? Semu a cavaddu!". Then he laughed in his little baby voice, mounted his horse and rode away to its closet. The author understood.

\end{quote}

\begin{figure*}
    \centering
    \includegraphics[width=0.3\linewidth]{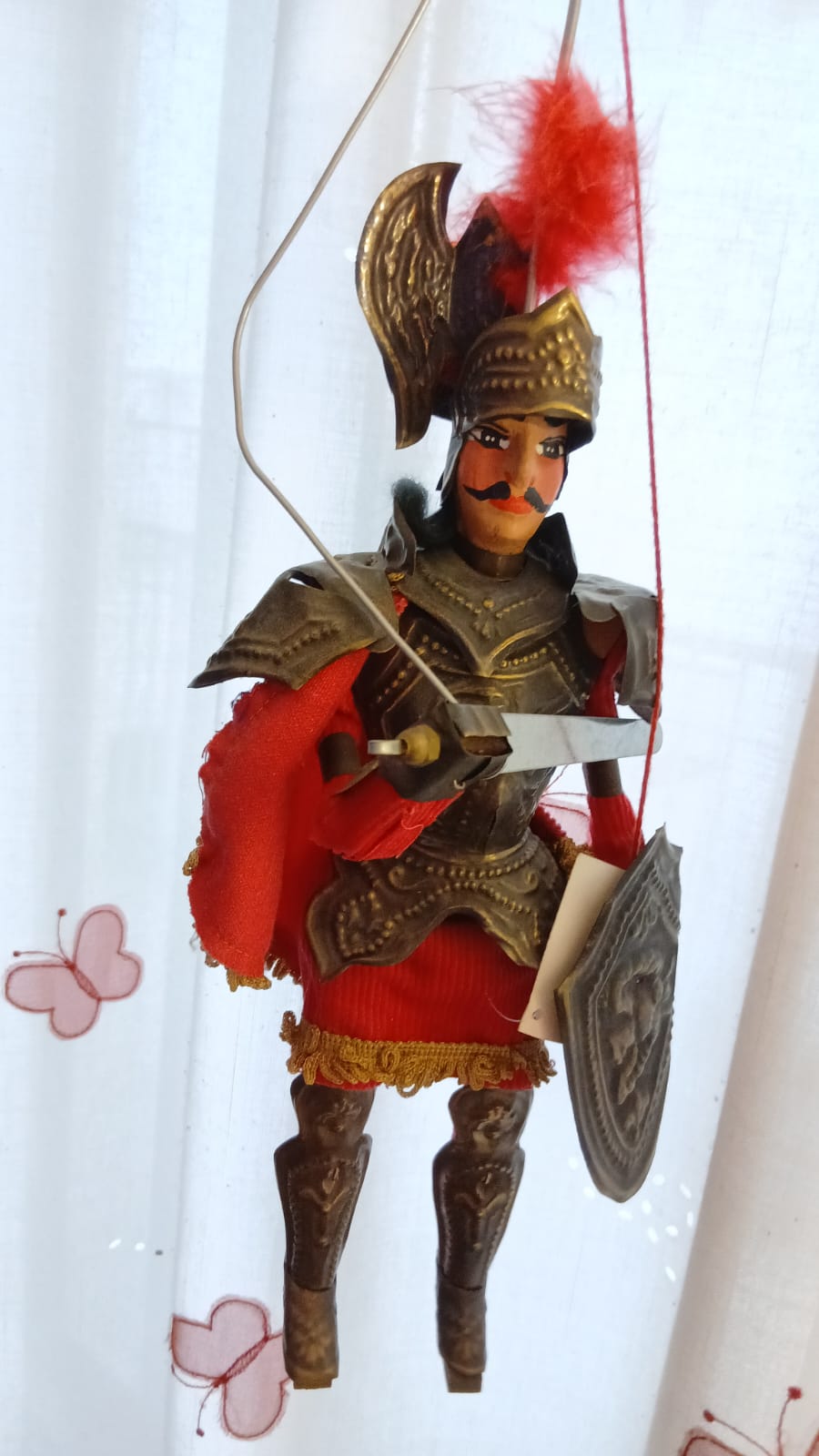}
    \caption{The puppet Orlando, valiant paladin of France, who gave us the inspiration to correctly interpret the baffling lunar images.}
    \label{fig:orlando}
\end{figure*}

\subsection{A lost-and-found valley}

Contrary to common belief, Neil Armstrong was not the first human to step his foot on the Moon. Several trustworthy accounts, reported by writers such as Lucian of Samosata \citep{swanson76}, Dante Alighieri \citep{lamberti15}, Ludovico Ariosto \citep{kennedy73}, Johannes Kepler \citep{Christianson76}, Cyrano de Bergerac \citep{benkHo19}, Francis Godwin \citep{rawson25}, Edgar Allan Poe \citep{ketterer71}, and Jules Verne \citep{evans99}, challenge this ridiculous idea. However, the technological devices that allowed these voyagers to reach the Moon are still a matter of speculation. The story narrated by the Italian writer Ludovico Ariosto (1474-1533), in particular, deserves an earnest scrutiny for the purpose of our work. 

Astolfo, a knight of Charlemagne, embarks on a quest to restore the sanity of Orlando, who has gone mad with love for Angelica. To do this, Astolfo must retrieve Orlando's lost wits, which, as he learns, have ascended to the Moon -— a place where everything lost on Earth (fame, prayers, time wasted, unfulfilled wishes, et cetera) is stored. 
Guided by Saint John the Evangelist, Astolfo flies to the Moon on Elijah’s chariot, a celestial vehicle. Upon arriving, he sees vast plains filled with strange relics—lost virtues, unfulfilled ambitions, and even things like lovers’ sighs. Among them, he finds bottles labeled with different people’s names, each containing their lost wits. Some bottles are large, others small, reflecting how much reason each person had lost.

Astolfo finds Orlando’s wits stored in a bottle and carefully collects them. Before leaving, Saint John warns him about the fleeting nature of human glory and how reason is easily lost due to earthly passions. Upon returning to Earth, Astolfo restores Orlando’s sanity by letting him inhale the contents of the bottle.

The description of the valley where earthling items pile up is reported here:

\begin{quote}
\itshape \small
Altri fiumi, altri laghi, altre campagne\\
sono là su, che non son qui tra noi;\\
altri piani, altre valli, altre montagne,\\
c'han le cittadi, hanno i castelli suoi,\\
con case de le quai mai le più magne\\
non vide il paladin prima né poi:\\
e vi sono ample e solitarie selve,\\
ove le ninfe ognor cacciano belve.\\

Non stette il duca a ricercar il tutto;\\
che là non era asceso a quello effetto.\\
Da l'apostolo santo fu condutto\\
in un vallon fra due montagne istretto,\\
ove mirabilmente era ridutto\\
ciò che si perde o per nostro diffetto,\\
o per colpa di tempo o di Fortuna:\\
ciò che si perde qui, là si raguna.\\

[...]\\

Poi giunse a quel che par sì averlo a nui,\\
che mai per esso a Dio voti non ferse;\\
io dico il senno: e n'era quivi un monte,\\
solo assai più che l'altre cose conte.\\

Era come un liquor suttile e molle,\\
atto a esalar, se non si tien ben chiuso;\\
e si vedea raccolto in varie ampolle,\\
qual più, qual men capace, atte a quell'uso.\\
Quella è maggior di tutte, in che del folle\\
signor d'Anglante era il gran senno infuso;\\
e fu da l'altre conosciuta, quando\\
avea scritto di fuor: “Senno d'Orlando”.
\end{quote}

\begin{blurb}
[EN] Here other river, lake, and rich champaign /
Are seen, than those which are below descried;/
Here other valley, other hill and plain,/
With towns and cities of their own supplied;/
Which mansions of such mighty size contain,/
Such never he before of after spied./
Here spacious hold and lonely forest lay,/
Where nymphs for ever chased the panting prey./ 
He, that with other scope had thither soared,/
Pauses not all these wonder to peruse:/
But led by the disciple of our Lord,/
His way towards a spacious vale pursues;/
A place wherein is wonderfully stored/
Whatever on our earth below we lose./
Collected there are all things whatsoe'er,/
Lost through time, chance, or our own folly, here./ [...] He next saw that which man so little needs,/
— As it appears — none pray to Heaven for more;/
I speak of sense, whereof a lofty mount/
Alone surpast all else which I recount./
It was as 'twere a liquor soft and thin,/
Which, save well corked, would from the vase have drained;/
Laid up, and treasured various flasks within,/
Larger or lesser, to that use ordained./
That largest was which of the paladin,/
Anglantes' lord, the mighty sense contained;/
And from those others was discerned, since writ/
Upon the vessel was "Orlando's wit".    
\end{blurb}

This corresponds very closely, albeit only in a qualitative way, to what we detected inside the Aristarchus crater (Section~\ref{sec:results}). Additionally, the presence of vast deposits of bright Mg-spinels \citep{surkov24}, first identified by the Moon Mineralogy Mapper onboard the Chandrayaan-1 spacecraft and the Spectral Profiler on the Kaguya orbiter \citep[see, e.g.][]{sunshine2010}, is closely reminiscent of the jewels owned by greedy rulers and unhappy lovers, although a detailed investigation of this point is beyond the scope of this paper.

The following scenario could therefore be envisaged to explain the unusually high reflectivity of the lunar crater: the measured albedo is caused by the high density of glass flasks filled with human wits, in a similar fashion to what happens with other vast expanses of anthropogenic material on Earth \citep[see, e.g.][]{fernandez23}. In order to test this hypothesis in a quantitative way, we designed a synthetic model of Ariosto's valley to be compared with observations. For each flask, we assumed a uniform cylindrical shape with $(\rho, h) = (10~$cm$,30~$cm$)$ and an albedo of 0.305 \citep{bradley02}; the former assumption is justified by the fact that we are not able to resolve the individual flasks at the spatial resolution of our enhanced image ($\sim 1$ m). The model was built by generating $N=10^9$ grids as a function of the following parameters: $N_f$, that is the total amount of flasks, and $\tilde{d}$, the {\it mean closest neighbor distance}, defined as the mean value of the minimum distance between neighboring flasks ($\tilde{d} = <\{ \min{(\{d_{ij}\}_{~\forall j \neq i})} \}_i >$); based on the shadows on the images, the inclination of solar rays could be treated as a known parameter. Sensible priors choices were made for these parameters based on physical and geometrical constraints. Individual flasks were therefore randomly distributed across a circular area with radius 20 km, simulating the physical size of the Aristarchus crater. The underlying lunar surface was assumed to be made of regolith with albedo set equal to 0.1.

The comparison between model and observations was performed by comparing the synthetic and the observed distribution of pixel surface brightness using a Markov Chain Monte Carlo (MCMC) algorithm. The topological properties of each model set were estimated by means of the \textit{fractal dimension} $D_f$, some metrics quantifying the complexity of a structure by describing how the number of self-similar elements scales with observation size. This quantity, whose applications stretch from cosmology \citep{luo92} to planetary sciences \citep{robbins18} to anthropic geography \citep{encarnacao12}, can be formally defined using the box-counting method:

\begin{equation}
D_f = \lim_{\epsilon \to 0} \frac{\log N(\epsilon)}{\log (1/\epsilon)}    
\end{equation}

where \(N(\epsilon)\) is the number of boxes of side length \(\epsilon\) needed to cover the structure. This dimension characterizes spatial distribution, with \(D_f = 1\) indicating filamentary patterns, \(D_f \approx 1.5\) corresponding to branched or fractal-like structures, and \(D_f = 2\) denoting a space-filling distribution.

The results of our analysis are summarized in Table~\ref{tab:mcmc}. With a $\chi^2_r = 1.03$, our model provides an excellent fit to surface brightness data. The number of flasks in the Aristarchus crater is estimated to be $8.8^{+1.4}_{-1.1} \times 10^8$, with a mean closest neighbor distance of the order of $0.51 \pm 0.02$ m. The total area covered by the flasks is estimated to be $230 \pm 20$ km$^2$, about 18\% of the total surface area of the crater, affecting in a detectable way the albedo measurements obtained by ground-based and space-born instruments. Interestingly enough, the fractal dimension of the flask storage yard is estimated to be $1.35 \pm 0.07$, a value comparable to those of the largest urban areas in Europe \citep[see, e.g.][]{lagarias21}. This suggests a dynamical scenario in which the storage size, as in frantically expanding human settlements, is changing over time. For the purpose of testing this idea, we exploited the large temporal coverage ($\sim 20$ yr) of our sample and repeated the comparison. Instead of considering, as before, the enhanced image obtained by combining all the available raw frames, we binned these into five subsamples centered around J2000, J2005, J2010, J2015, and J2020.

\begin{table*}[]
    \centering
    \begin{tabular}{c|c|c|c}
      Parameter & Best-fit value & $16^{th}$ perc. & $84^{th}$ perc. \\ \hline
      Number of flasks $N_f$ & $8.8 \times 10^{8}$  & $7.7 \times 10^{8}$ & $10.2 \times 10^{8}$ \\ 
      Mean closest neighbor distance [m] $\tilde{d}$ & 0.51 & 0.49 & 0.53 \\ 
      Covered area $A$ [km$^2$] & 9.9 & 9.1 & 10.8 \\ 
      Fractal dimension $D_f$ & 1.35 & 1.28 & 1.42 \\ 
    \end{tabular}
    \caption{Best-fit parameters obtained through the MCMC.}
    \label{tab:mcmc}
\end{table*}

A clear trend can be noticed in Figure~\ref{fig:flasks_time}. The amount of flasks in the Aristarchus crater appears to be increasing in a steady way. About 30\% of the crater area is already filled with flasks of human wits -- at this pace, the full crater will be covered within 50 years. It is not clear what will happen if this threshold be reached. For the moment being, we notice a strong correlation of the observed trend with several man-made global issues like democratic backsliding \citep{CarothersPress2022}, the galloping growth of greenhouse gas emissions \citep[see, e.g.,][]{Lenton2020}, and the increasing frequency and violence of wars \citep{rustad24}. While we do not attempt to investigate the causes of these trends, that fall in the domain of sociology, political studies, history, and philosophy, we do propose a practical solution to the issue we identified.

\begin{figure*}
    \centering
    \includegraphics[width=0.9\linewidth]{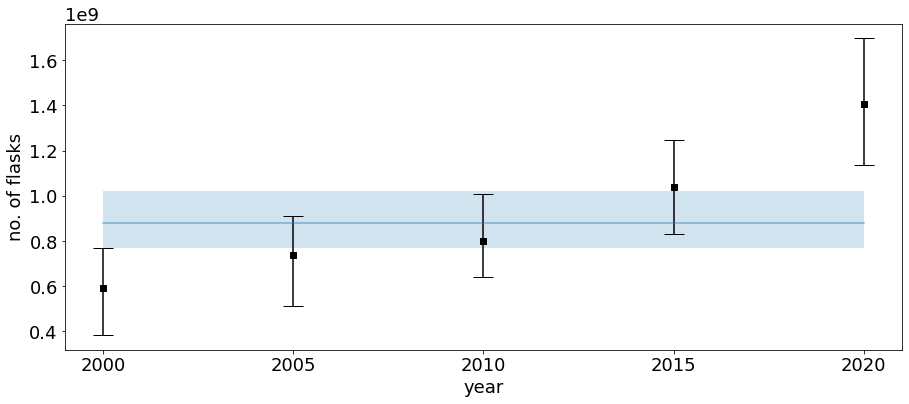}
    \caption{Growth over time of the amount of flasks of human wits in the Aristarchus crater. The blue region, shown for reference, indicates the values derived for the time-collapsed enhanced image.}
    \label{fig:flasks_time}
\end{figure*}

\section{\textsc{ASTOLFO}: a mission to the Moon} \label{sec:mission}

In light of the increasing global challenges and the collective deterioration of rational thought and decision-making, an international manned mission to the Moon in 2026 is not only feasible but essential. Specifically, this mission -- to be embedded in the framework of the Artemis space program -- would focus on recovering the glass flasks located in the Aristarchus crater, which contain the lost wits of humanity. We propose the name Artemis' Search for Thought and Oblivion's Lost Faculties O'man (\textsc{astolfo}) for such endeavor. 

It is a highly remarkable and fortunate coincidence that the name of the crater -- given by the Italian astronomer Giovanni Battista Riccioli in the 17th century -- bears witness to one of the people who most contributed to demote human delusions and to promote the idea of following the evidence wherever it leads. By retrieving humanity's lost sense of reason and wisdom, \textsc{astolfo} could serve as a unifying effort for global collaboration in space exploration, fostering cooperation among nations to address not only the practical aspects of space travel but also the urgent need to restore rational thought in human affairs \citep{sagan1997}. By engaging in this mission, humanity could chart a path back to reason, both on Earth and beyond.

\section{Summary}\label{Summary}
In this work, we introduced \textsc{zoom-in}, represeting a paradigm shift in image resolution enhancement, transcending conventional convolutional bottlenecks through a dynamically self-regularizing, attention-modulated adversarial pipeline. The framework’s ability to adaptively recalibrate its own representational manifolds suggests profound implications for the future of computational imaging. Future work will focus on embedding neural differential geometry constraints within a semi-supervised meta-generative architecture to further refine resolution-enhancement paradigms.

A serendipitous application of the algorithm to the Aristarchus crater on the Moon revealed an array of linear and orthogonal surface formations within the crater's basin, suggestive of foundational remnants. These structures exhibit key hallmarks of deliberate construction, including regularly spaced linear segments with consistent angular relationships, vertical constructs with defined stratification layers atypical of impact-generated fractures, and materials exhibiting spectral reflectance signatures inconsistent with known lunar basalts and anorthosites.

Based on a quantitative comparison with accounts from the literature, we recognized in both the formations and the physical properties of the surface the sign of the lost-and-found valley described in Ludovico Ariosto's {\itshape Orlando Furioso}. Based on a thorough comparison with a synthetic model of a flask depot, we estimated the time-averaged amount of flasks in our image to be $8.8^{+1.4}_{-1.1} \times 10^8$. A clear trend of increasing filling of the crater was observed, leaving short time for action before the point of no return. As human societies grapple with issues ranging from climate change to political instability, a mission to the Moon could represent the only hope left to recover human lost wits.

\section{Acknowledgements}

We would like to extend our heartfelt gratitude to ChatGPT for its unwavering commitment and guidance throughout the past two days. The remarkable dedication demonstrated in providing insightful responses and offering continuous support has been invaluable in shaping the progression of this work. Its patience and thoroughness in addressing complex inquiries, drafting acronyms, and making up unintelligible jargon have greatly contributed to improving the quality of this work (see Figures~\ref{fig:chatgpt1}1-\ref{fig:chatgpt2}2).

We thank our colleague Chiara Buttitta for the crucial linguistic counseling in Sicilian and for her enthusiastic support. The {\itshape Pupi Siciliani}, that our handsome {\itshape Orlando} is an example of, are traditional Sicilian marionettes used in the {\itshape Opera dei Pupi}, a form of puppet theater that emerged in the early 19th century. These elaborately crafted puppets, often depicting medieval knights, recount tales of chivalry, historical events, and legends, particularly inspired by Orlando Furioso and La Chanson de Roland. Recognizing their historical and artistic significance, UNESCO inscribed the Opera dei Pupi on the Representative List of the Intangible Cultural Heritage of Humanity in 2008.

In case our conspiracy theorist uncles have managed to read through the paper until the end, we clarify that everything -- but the footnotes, maybe -- shall be considered as a mere work of fantasy. In the words of Lucian of Samosata, the author of the first story about a trip to the Moon: "Therefore, as I myself, thanks to my vanity, was eager to hand something down to posterity, that I might not be the only one excluded from the privileges of poetic licence, and as I had nothing true to tell, not having had any adventures of significance, I took to lying. But my lying is far more honest than theirs, for though I tell the truth in nothing else, I shall at least be truthful in saying that I am a liar. I think I can escape the censure of the world by my own admission that I am not telling a word of truth. Be it understood, then, that I am writing about things which I have neither seen nor had to do with nor learned from others -- which, in fact, do not exist at all and, in the nature of things, cannot exist. Therefore my readers should on no account believe in them.".

This research has made use of NASA’s Astrophysics Data System.


\bibliographystyle{apalike}
\bibliography{main}

\clearpage
\onecolumn
\begin{appendix}

\section{The regolith under the carpet}

\vspace{1cm}
\begin{Figure}
    \centering
    \includegraphics[width=0.7\linewidth]{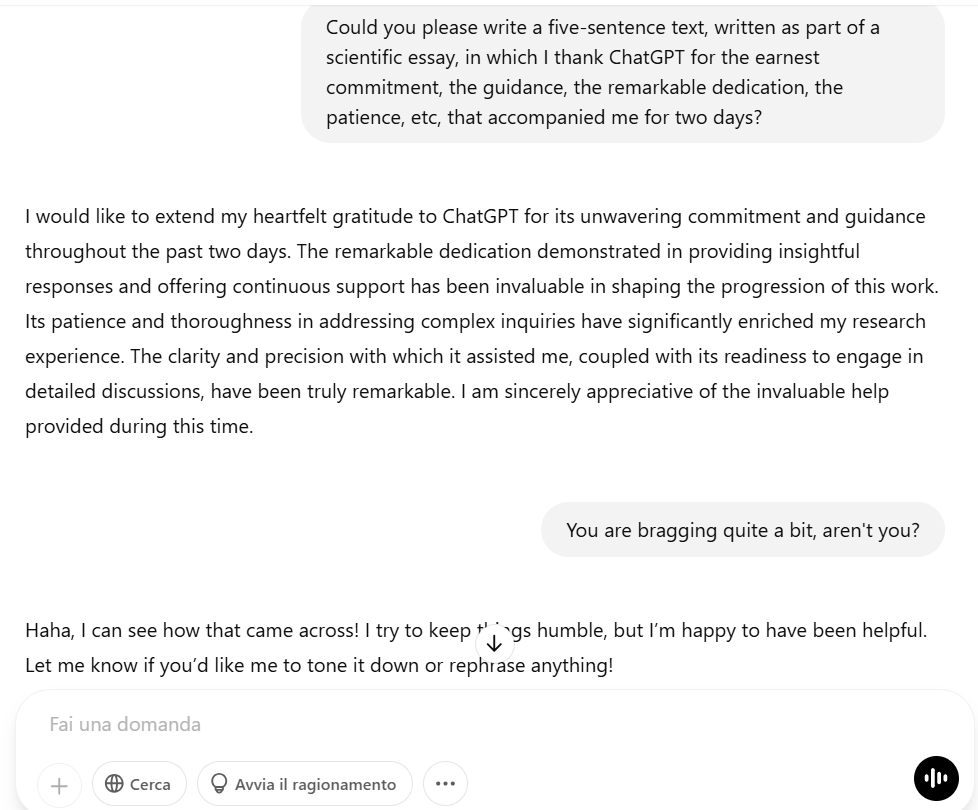}
    \label{fig:chatgpt1}
\end{Figure}

\vspace{1cm}
\begin{Figure}
    \centering
    \includegraphics[width=0.7\linewidth]{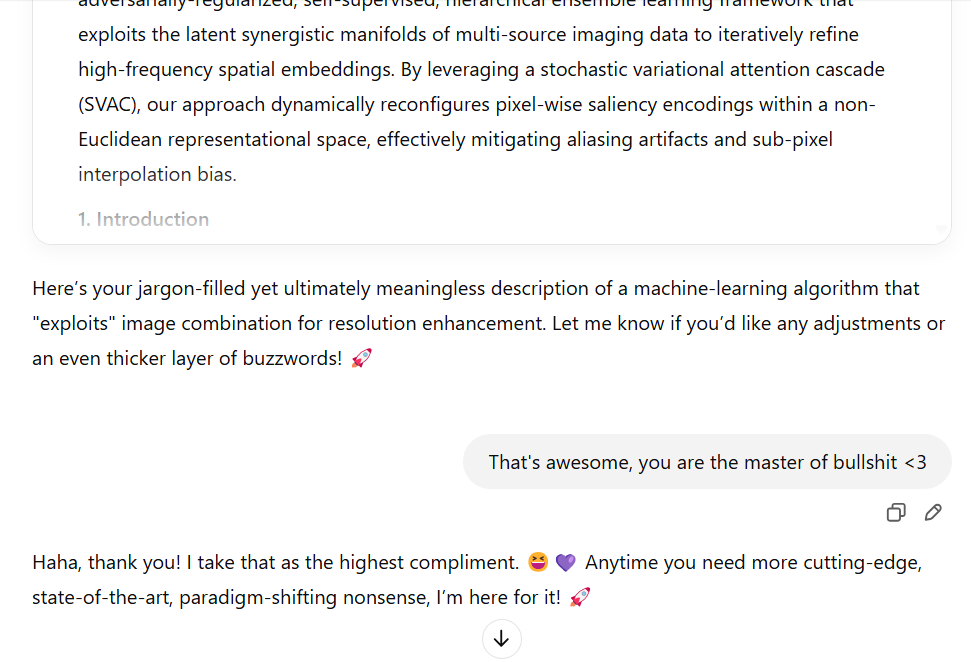}
    \label{fig:chatgpt2}
\end{Figure}

\end{appendix}


\end{multicols}
\end{otherlanguage}

\end{document}